\documentstyle[preprint,aps,epsf]{revtex}
\begin{document}
\draft
\preprint{\vbox{%\hbox{KIAS-P99020}
      \hbox{hep-ph/9903395}}}
\title{Electroweak Symmetry Breaking due to Confinement}
\author{Chun Liu}
\vspace{0.5cm}
\address{Korea Institute for Advanced Study\\ 
207-43 Cheongryangri-dong, Dongdaemun-gu, Seoul 130-012, Korea\\
liuc@kias.re.kr}
\maketitle
\thispagestyle{empty} 
\setcounter{page}{1}
\begin{abstract}
Within the framework of gauge mediated supersymmetry breaking, we 
consider an electroweak symmetry breaking pattern in which there is no
conventional $\mu$ term.  The pattern is made appealing through realizing 
it as low energy effective description of a supersymmetric Yang-Mills
theory which is of confinement.  Phenomenological implications are
discussed.\\


%Keywords: electroweak symmetry breaking, supersymmetric gauge theory, 
%confinement.
\end{abstract}

\newpage

%%%%%%%%%%%%
%\section{Introduction}
%\label{sec:introduction}
%%%%%%%%%%%%

Supersymmetry \cite{susy} provides a solution to the gauge hierarchy
problem if it breaks dynamically \cite{witten}.  It has been realized 
that the breaking should occur in a hidden sector, which is then 
communicated to the observable sector.  In this paper, we consider the 
scenario in which the communication is via gauge 
interactions \cite{dine1,dine2}.  Generally, the gauge mediated 
supersymmetry breaking (GMSB) models have a so-called $\mu$ 
problem \cite{dine2,mu1}, namely either the $\mu$ term is at the weak 
scale and the $B\mu$ term is unnaturally large, or $B\mu$ is at the weak
scale and $\mu$ is very small.  That means, there are some difficulties in 
getting right electro-weak symmetry breaking (EWSB).  Although several ways 
were suggested for solving this problem \cite{dine2,mu1,mu2}, it would 
be desirable to find more simple solution.

Instead of generating the $\mu$ term, we suggest to study the EWSB by the 
following superpotential,
\begin{equation}
W_{EWSB} = -\lambda X (H_u H_d - \mu^2) \,,
\label{1}
\end{equation}
where $H_u$ and $H_d$ are the two Higgs doublets, $X$ a standard model
singlet; $\mu$ is the EWSB scale, and $\lambda$
the coupling constant.  The physical implications of above superpotential 
will be discussed later.  In fact, it was used in early stage of the 
supersymmetry phenomenology \cite{review}.  Spontaneous EWSB is obtained as 
\begin{equation}
v_u = v_d = \mu \,,
\label{2}
\end{equation}
where $v_u$ and $v_d$ denote vacuum expectation values (vevs) of the doublet 
Higgs fields, and the other fields have vanishing vevs.  Note that $W_{EWSB}$ 
does not break supersymmetry.  After taking relevant soft masses into 
consideration, it can be seen that there are no light Higgs and light 
Higgsino particles.

%%%%%%%%%%%
%\section{Model}
%\label{sec:model}
%%%%%%%%%%%

The superpotential $W_{EWSB}$ of Eq. (\ref{1}) may have fundamental reasons.  
It can be an effective theory of a more fundamental theory.  The results of 
supersymmetric Yang-Mills theory \cite{seiberg} can be used to realize this 
idea.  To be specific, we exploit a model of Intriligator, Seiberg and 
Shenker \cite{iss}.  Introduce a supersymmetric SU(2) gauge interaction with 
a single matter superfield $Q$ in the $I=3/2$ representation.  This theory 
is believed to be of confinement.  The basic gauge singlet field is $u=Q^4$ 
with a totally symmetric contraction of the gauge indices.  The quantum 
theory has a moduli space of degenerate vacua labeled by the vev of $u$.  
The nontrivial check of the 't Hooft anomaly matching conditions implies that 
the K\"ahler potential at low energy is
\begin{equation}
K \sim u^{\dag} u |\Lambda|^{-6} ~~~ {\rm for} ~~~ u^{\dag} u < \Lambda^8 \,,
\label{3}
\end{equation}
with $\Lambda$ being the dynamical scale of the SU(2) interaction.  Perturbing 
the theory by a tree level superpotential 
$\displaystyle\frac{k}{m}u$ with $m$ being some 
new physics scale and $k$ the dimensionless 
coupling coefficient would break supersymmetry.  
To achieve EWSB other than supersymmetry breaking, we assume that the new
physics further couples $Q$ 
with standard model Higgs fields which are singlet under this SU(2).  The 
low energy effective superpotential is written as 
\begin{equation}
W_{eff} = \lambda_1 m H_uH_d+\frac{k}{m}u-\frac{c}{m^3}uH_uH_d \,,
\label{4}
\end{equation}
where $\lambda_1$ and $c$ are dimensionless coupling constants.  
By field redefinition $\displaystyle u\to u+m^4\frac{\lambda_1}{c}$, 
$W_{eff}$ becomes to 
\begin{equation}
W_{eff} = \frac{k}{m}u-\frac{c}{m^3}uH_uH_d \,,
\label{4a}
\end{equation}
plus some unphysical constant, where we denote the redefined field still as
$u$ without confusing.  
Note that the K\"ahler potential does not change under this 
redefinition.  To the order of $1/m^3$, the general effective superpotential
includes terms $\lambda_2(H_uH_d)^2/m+\lambda_3(H_uH_d)^3/m^3$ with 
$\lambda_2$ and $\lambda_3$ being dimensionless constants.  The presence
of these terms does not modify the above-discussed EWSB qualitatively.  
The point is that whenever there appears a term proportional to $H_uH_d$,
it can be removed through the above procedure of field redefinition.
In addition, $\lambda_2$ and $\lambda_3$ can be small.  The smallness is 
natural in the sense of 't Hooft due to the non-renormalization theorem in 
supersymmetry.
Besides that $u$ is a composite field
with dimension $4$, Eqs. (\ref{3}) and (\ref{4a}) is the same as the physics 
by Eq. (\ref{1}) with an elementary $X$.  

Quantitatively, rescaled field $u/\Lambda^3$ corresponds to $X$, and then 
\begin{equation}
\lambda = c\frac{\Lambda^3}{m^3} \,, ~~~~~ \mu^2 = \frac{k}{c}m^2 \,.
\label{5}
\end{equation}
It can be seen from Eq. (\ref{1}) that keeping Higgsino mass at weak scale 
requires $\lambda\sim O(1)$.  So numerically $c$ is $\sim m^3/\Lambda^3 > 1$.  
This is consistent with $\mu^2 < m^2$ if $k$ is $O(1)$, 
$\mu^2\sim k\displaystyle\frac{\Lambda}{m}\Lambda^2$.  By taking $\Lambda$ to 
be ($100-1000$) GeV, $m$ should be about ($10^2-10^5$) GeV.  Therefore, 
viable EWSB can indeed occur dynamically due to confinement of a 
supersymmetric gauge theory with certain effective tree level superpotential.

It is necessary to discuss theoretical implications of the above described 
EWSB.  First, the breaking scale $\mu$ is not generated by the supersymmetry 
breaking which has not been dealt with yet.  It is related to the SU(2) 
dynamical scale $\Lambda$ and the new physics scale $m$.  However, the EWSB 
is still tied to supersymmetry itself.  Supersymmetry is necessary to keep 
the gauge hierarchy whenever there are elementary scalar particles.  Second, 
it is not radiative breaking.  Once new scales are introduced for generating 
the scale $\mu$, radiative breaking is no longer a requirement of 
simplicity.  It is natural to relate the scales $\Lambda$ and $m$  to the 
EWSB directly.  Third, we wonder if there is a relation between the scale 
$m$ and the supersymmetry breaking scale.  For instance, the scale $m$ can 
be at $10^4$ GeV which might also be the supersymmetry breaking scale.  It 
would be interesting that the EWSB is finally connected to the supersymmetry 
breaking.  Fourth, in principle this EWSB mechanism may also apply to the
case of supergravity.  Of course, it seems to have less relation with the 
supersymmetry breaking in this case which is therefore less interesting.

Supersymmetry breaks dynamically in another sector.  There are several ways 
to get the breaking \cite{review2}.  For simplicity, we can still adopt the 
model of Ref. \cite{iss}.  Introduce another SU(2) with single matter $Q'$ 
in the $I=3/2$ representation.  The singlet composite field is $u'=Q'^4$.  
The tree level superpotential $\displaystyle\frac{u'}{m'}$ with $m'$ being 
some scale breaks supersymmetry dynamically.  To mediate the supersymmetry 
breaking to the standard model sector, we introduce the so-called messenger 
fields which are singlet under this new SU(2) but in the vector representation 
under the standard model gauge group.  Couple $u'$ to the messengers in the 
way like the $u$ field to the Higgs fields in Eq. (\ref{4}).  The difference 
here is that the messenger mass terms in the superpotential cannot be removed
by field redefinition so as to avoid 
the messengers developing vevs.  
And these mass terms break R-symmetry explicitly \cite{dine1}.  
Supersymmetry breaking is mediated to the 
standard model sector through loops.  In fact, the effective theory obtained 
from the above is just the O'Raifeartaigh model \cite{tree} used by Dine and 
Fischler in Ref. \cite{dine1} which gives details of the messenger content.

%%%%%%%%%%%
%\section{Phenomenology}
%\label{sec:pheno}
%%%%%%%%%%%

The phenomenological implications of the EWSB described in this paper should
be stressed.  The Higgs vevs are determined by the superpotential 
Eq. (\ref{1}), the supersymmetric standard model gauge interactions and the 
soft masses,
\begin{equation}
V = |\lambda(v_uv_d-\mu^2)|^2+\frac{1}{8}(g^2+g'^2)(v_u^2-v_d^2)^2
+M^2v_u^2+M^2v_d^2 \,,
\label{6}
\end{equation}
where $g$ and $g'$ are the standard model SU(2)$\times$U(1) gauge coupling 
constants, $M$ the soft mass of the Higgs particles.  The minimum of $V$ 
results in Eq. (\ref{2}).  Hence
\begin{equation}
\tan\beta \equiv \frac{v_u}{v_d} = 1 \,.
\label{7}
\end{equation}
Note that the usual phenomenological constraints on $\tan\beta$ in the 
minimal supersymmetric standard model (MSSM) do not apply here, because the 
EWSB is not radiative breaking.  Compared with the particle spectra of the 
MSSM, there is one more neutral Higgs and one more neutralino because of the 
introduction of $X$ field.  Due to the tree level electroweak breaking and 
the additional coupling $\lambda$, the spectra of the scalar bosons and the 
neutralinos are less constrained.  Nevertheless they are all around the weak 
scale.  Let us look at the neutralino masses which are given as
\begin{equation}
(\tilde{\phi}^0_d~~ \tilde{\phi}^0_u~~ \tilde{W}^3~~ \tilde{B}~~ \tilde{X})
\left(\begin{array}{ccccc}
0              &0             &gv_d/\sqrt{2} &-g'v_d/\sqrt{2} &\lambda v_u\\
0              &0             &-gv_u/\sqrt{2}&g'v_u/\sqrt{2}  &\lambda v_d\\
gv_d/\sqrt{2}  &-gv_u/\sqrt{2}&M_{\tilde{W}} &0               &0          \\
-g'v_d/\sqrt{2}&g'v_u/\sqrt{2}&0             &M_{\tilde{B}}   &0          \\
\lambda v_u    &\lambda v_d   &0             &0               &0
\end{array}
\right)
\left(\begin{array}{c}
\tilde{\phi}^0_d\\ \tilde{\phi}^0_u \\ \tilde{W}^3 \\ \tilde{B}\\ \tilde{X}
\end{array}
\right)\,,
\label{8}
\end{equation}
where $\tilde{\phi}_d$, $\tilde{\phi}_u$ and $\tilde{X}$ stand for the 
fermion components of $H_d$, $H_u$ and $X$.  $\tilde{W}$ and $\tilde{B}$ are 
Wino and Bino with soft masses $M_{\tilde{W}}$ and $M_{\tilde{B}}$ 
respectively.  The determinant of the matrix is about $M_Z^4M_{\tilde{W}}$.  
We see explicitly that there is no light Higgsino.  And all the neutralinos 
are around the weak scale.  The chargino mass matrix is more predictive, 
\begin{equation}
(\tilde{\phi}^+_u~~ \tilde{W}^+)
\left(\begin{array}{cc}
0              &M_W          \\
M_W            &M_{\tilde{W}}
\end{array}
\right)
\left(\begin{array}{c}
\tilde{\phi}^-_u \\ \tilde{W}^-
\end{array}
\right)\,.
\label{9}
\end{equation}
Because of the absence of conventional $\mu$ term, the two chargino mass
product satisfies: $M_{\tilde{\chi}_1^{\pm}}M_{\tilde{\chi}_2^{\pm}}=M_W^2$.  
$M_{\tilde{W}}\neq 0$ leads to that one of the charginos must be lighter 
than the W boson.  Such a chargino is within the experimental reach.  
However, if the lightest neutralino mass is close to this chargino mass 
within a few GeV, it is hard to be detected.  

%%%%%%%%%%%
%\section{Summary}
%\label{sec:summary}
%%%%%%%%%%%

In summary, an old EWSB pattern has been re-suggested to avoid the $\mu$ 
problem in the GMSB scenario.  Our main point is that it can be effective 
description of a more fundamental supersymmetric gauge theory which is of 
confinement.  Phenomenologically, additional neutral Higgs and one more 
neutralino are predicted with masses around the weak scale.  One of the 
charginos is lighter than the W gauge boson.

Several remarks should be made finally.  (i) This model is originally 
motivated 
by the works of Ref. \cite{flavor} which aim at the flavor problem.  
Slight lepton number violation can be introduced into the model.  The scalar
neutrinos develop small vevs.  In this case the $\tan\beta$ deviates from 
unity slightly.
(ii)  
The relation between EWSB and super-Yang-Mills theory is not unique.  In a 
recent model of Ref. \cite{mu3}, conventional $\mu$ term is generated 
dynamically.  
(iii) This EWSB mechanism is similar to the spirit of the technicolor 
\cite{tc}.  The electroweak symmetry breaking is triggered by a strong SU(2)
interaction.  However, there is distinction, that is the property
of the strong gauge interaction used in this mechanism is not spontaneous
chiral symmetry breaking, but confinement.

%%%%%%%%%%%%%%%%%%%%%%%%%%%%%%%%%%%%%%%%%%%%%%%%
\acknowledgments
%%%%%%%%%%%%%%%%%%%%%%%%%%%

I would like to thank E.J. Chun and S.Y. Choi for helpful discussions.

%\newpage

\end{document}